\documentclass[review]{article}

\usepackage{amssymb, amsthm,amsmath, bm, bbm, mathrsfs, graphicx, color}
\usepackage[]{algorithm2e}
\usepackage{algorithmic}
\usepackage{mdframed}
\usepackage{setspace}
\usepackage{latexsym}
\usepackage{kpfonts}
\usepackage{multicol}
\usepackage{natbib}
\usepackage{hyperref}
\usepackage{perpage} 
\MakePerPage{footnote}

\usepackage{setspace}
\usepackage{setspace}

\begin{document}
\onehalfspacing
\begin{center}
\textbf{\sc \Large  Rbox: an integrated R package for ATOM Editor }

\vspace{0.3cm}

{Saeid Amiri\footnote{ saeid.amiri1@gmail.com\\ https://atom.io/packages/rbox}}

{\it  Department of Natural and Applied Sciences,  University of Wisconsin-Green Bay, Green Bay, WI, USA  }
\end{center}


\begin{abstract}
R is a  programming language and environment that is a central tool in the applied sciences for writing program.  Its impact
on the development  of modern statistics is inevitable.  Current research, especially for big data may not be done solely using R and will likely use different programming languages; hence, having a modern
integrated development environment (IDE) is very important. Atom editor is
modern IDE that is developed by GitHub, it  is described as "A hackable text editor for the 21st Century". This report is intended to present a package deployed entitled
\textit{Rbox} that allows Atom Editor to write and run codes professionally  in R.
\end{abstract}
 

\section{Introduction}
\noindent Because of computational advances, technological modernization,
substantial cost-reductions, and the broad proliferation of modern cloud services, 
the use of data has rapidly increased over the past decade. The explosion of data (size, complexity, heterogeneity,
scale, and incongruence) parallels enormous methodological, analytical
and visualization developments. For instance, big data can serve as
a powerful proxy for various observed phenotypes and diagnostic traits;
enable the forecasting, modeling,  and prediction of disease prevalence; 
and impact our understanding of human health and disease. Typically,
such large amount of biomedical and health data are high-dimensional, and finding
underlying patterns, associations, latent relations, causal effects,
etc. is challenging. In addition, the computational burden of processing
complex multi-source data is heavy and often leads degenerative-statistics,
violations of parametric assumptions, lack of convergence, and ultimately
biased scientific inference. New efficient, and reliable techniques
are necessary to cope with the increase in data volume and complexity.
Having highly efficient programing tools for carrying out advanced statistical projects is essential.

R is a popular programming language platform used by many researchers
and scientists  because of its functionality, reliability, scalability,
open-sourceness, and crowd source support. R is the \textit{de facto}
standard for data analysis; anyone can download, view, contribute and
expand codes, protocols or scripts via its infrastructure CRAN. R is helping to develop the modern movement in data science, because statistics researchers with little programming knowledge can easily write codes and evaluate their theories numerically. Accompanying code supplements with the paper is an efficient method of introducing the novelty of
methods to other disciplines, see \citet{gentleman2007statistical}.
Rstudio impacts the attractiveness using R, and developing new libraries, 
but it lacks in-line running code. In-line coding allows
the effective exploration and rerunning of code. Jupyter is a set of open-source
software tools for interactive and exploratory computing.  Its main
application is the Jupyter Notebook, a web-based interactive computing
platform that interweaves written analysis and coding, it allows users
to author computational narratives that combine live code, equations,
narrative text, interactive user interfaces, and other rich media see
\citet{perez2015project}, which make it very useful for collaborating on
the numerical aspects of research. It is also used with tremendous frequency among
code educator.   The Jupyter command shell used with R is called IPython. It  not only provides
an enhanced shell, but also facilitates parallel computing in Python
and  accessibility as an interactive computation environment.
The IRKernel is R's counterpart to IPython and provides an interactive
environment for scientific computing, built with the purpose of developing
an interactive environment for accessing R for scientific computing\footnote{https://jupyter.org/}.

With the current developments in data sciences, coding in several
programming languages is inevitable,  and  programming can not  relay solely on R.
For instance Python is
a core language in machine learning development,
also received attention from statisticians.  There are several IDE and one, \textit{Rodeo} looks similar Rstudio, although it is not comparable
to Atom editor. When working on a project, finding an editor with access to several
programming languages is desirable. Atom editor is a free and modern IDE
developed by the GitHub community and is used as an editor for coding, debugging, and managing projects in several programming
languages,  including R and Python. Atom editor is equally responsive
to users' needs for both writing and running codes, and it is quickly making its way as a new modern editor that supports a number of programming languages, along with other code editors; such as Visual Studio Code, Sublime, and vim.  

The goal of this paper is to present a package  entitled \textit{Rbox}
designed for running R in Atom editor. Before focusing on Atom editor, it is beneficial
to discuss why R will remain among the top  programs used for working with
data.  It has very advanced libraries for running the parallel computations, 
which can be used for computer-intensive methods and working with big data, see Section \ref{parallel}. The use of R codes
in web-based interfaces has not been widely adopted, this possibility is discussed in Section \ref{webinter}. Section\ref{interact} outlines  R's ability to interact with other software.  Section \ref{atomide} presents \textit{Rbox} and discusses Atom editor.

\section{Parallel calculation}
\label{parallel} Scalability in running an algorithm is essential. 
R has the ability to run calculations in parallel, which is
important for complicated and advanced computation.  For instance,  in the
statistical texts, the ensembling has received much attention. The idea
of ensembles is to assume that there is a fixed large model, for which the
modeling (prediction) can be done by pooling the results from carefully
chosen smaller models. The final model can be found by a consensus
value of the submodels and can provide  better performance than any one of
the components used to form it see \cite{amiri2017clustering}. The ability
to run the ensemble calculations  in parallel is essential; otherwise,  they might
not be scalable. Modern PCs and laptops have multi-core processors  that allow the  computation to be done in parallel.   Several libraries in R implemented  to run computations in parallel, for instance,  
see \texttt{parallel}, \texttt{foreach}, \texttt{doParallel}.  More details about the packages developed for parallel computation can be found on parallelr's website \footnote{http://www.parallelr.com/}. 
Parallelizing codes in R is very simple and does not require advanced programming skills.

It may be impossible to work with big data without using parallel computation, cluster computers, and supercomputer.  Running parallel computations in R  using cluster computers and supercomputers using Linux and Unix-like kernels, for instance Slurm, SGE,
and LSF,  is very easy and helps statisticians with little knowledge of programming run big projects. 



\section{Web-based interface}
\label{webinter} Some of the released functions
in the R libraries might not be user-friendly, and it may be quite overwhelming
for  users to grasp and correctly use all methods and packages.
Practitioners are expected to be familiar with the manual, assumptions,
and syntax of R, which may be time consuming and challenging for interdisciplinary
researchers. To overcome these problems, many developers have begun
to build and share graphical user interfaces that lower the amount of  initial
knowledge required to use R. This goal motivated researchers to design, deploy, and support a web-based interface that users can quickly, efficiently
and effectively run without the need for a complex R computational
platform. Here we review a number of services, interfaces, and APIs prior
to developing the web interface.
\begin{itemize}
\item Shiny: Rstudio has developed a web application framework for R that is
a simple way to turn  Rs codes into a web-based console.  The website
version in Shiny's server is not integrated with non-Shiny
clients; the free account has limitations    
on running times and on the number of
websites that can be built, and allocates memory inefficiently.
As \citet{ooms2014opencpu} discussed, Shiny lacks interoperability.
Because it is, in essence, a remote R console, it  does not specify any
standardized interface for calling methods, data I/O, etc.
\item OpenCPU,  see\citet{ooms2013opencpu}, is a system for turning an R package
into a deployable web application that is based on JavaScript. It
can be run from GitHub, which is a free repository for open-source projects.
This development motivated efforts to design, deploy, and support a web-based
interface that users can quickly, efficiently, and effectively run
without the need for a complex R computational platform. 
\end{itemize}

\section{Interactivity with other software}
\label{interact}
R is a shell script language that is designed to be run by a command-line
interpreter. This property helps make R interactive and accessible
software.  Many of the codes behind R are written in Fortran and C. Several
libraries in R are packed to handle codes from other software; for
instance, \texttt{rPython} runs Python codes from R. R has very powerful
tools for generate advanced and unique graphs on which users can work interactively; the graphical ability of R is well-known, and many libraries
in R achieve amazing graphical presentations. Researchers
develop advanced graphs, and while deploying them using R's technology
might not be trivial, R works interactively with other technology. 
For instance, advanced graphs are developed in Plotly (an
open source JavaScript graphing library [plotly.js]), and an
R package entitled "Plotly"\footnote{https://plot.ly/r/}
 can create interactive web graphics.

Having a shell script helps make R favorite piece of software for running  code
inside other software; for instance, Azure Machine Learning Studio  programs with R and Python. R has its own kernel that allows it to be run inside other software.

\section{Modern IDE}
\label{atomide} 
Rstudio provides an advanced editor for R with several tools, especially for deploying libraries.  One of the downsides of Rstudio is that it is designed solely for R, and working with several programming languages is inevitable. 
Therefore, having modern editors with access to advanced programming languages and coding abilities is essential.

Atom editor is actually a web application that looks like a regular app. 
It is based on  JavaScript; hence, its user interface can be changed with a few
lines of codes and is highly configurable, and the  user can take control
of  the editor. The model for the development of  Atom editor is the same as  that for R; it has a core that is powered by GitHub's team,  and packages
are added to develop the  editor for different applications; it has a repository
\footnote{https://atom.io/packages}; and developers can deploy and
update packages. 
 Because Atom editor has very modern tools that can be used by statisticians
and that outperform Rstudio in exploring codes, the author of this paper collected
a toolbox, entitled "Rbox"  for running R under Atom editor.  It can be downloaded directly from its website \footnote{atom.io}. Once the main software is installed, the necessary packages can be installed using the shell commands \textit{apm install packagename} or  Atom's package installer;  Atom$>$ Preferences$>$ Install (in Win: File$>$ Settings$>$ Install).

To work with R in Atom editor, install Hydrogen \footnote{https://atom.io/packages/Hydrogen},
termination \footnote{https://atom.io/packages/termination}, Rbox
\footnote{https://atom.io/packages/rbox}. Rbox defines the R grammar
in Atom editor to work and facilitate with snippets. Atom editor is one of
many IDEs aimed at providing a powerful and modern environment with access to autocompletions. A package entitled "Autocompletion-R"  with access to helps  \footnote{http://stat.ethz.ch/R-manual/R-patched/library/} is available.
The user can add own snippets to expedite codding. Autocompletion is very
useful when the user  is working with different programming languages.  To
accept the suggestion provided by Atom press the tab \textbackslash{}enter key.  The down arrow key can be used to select  the right option.

To write script, R must be chosen from the Grammar in the editor using
\textit{Shift+Ctrl+L} or by clicking the current grammar name in the
status bar,  Atom editor then knows the coding  language and accesses the snippets
available in \textit{Rbox} for R. The scripts can be run
in the terminal without leaving Atom editor, as the editor accesses to the terminal,
and the code can be transferred to it. This can be done using clicking on the plus sign on the
bottom left  to open a terminal (\textit{Shift+Cmd+T} in macOS or \textit{Shift+Alt+T} in Win). The terminal will access R when R is typed. To transfer the code to the terminal, select code and press \textit{Shift+Alt+R}
in macOS (or \textit{Alt+R} in Win), the code can also be executed line by
line as well. Several terminals can be opened at once.  Atom editor Grammar has access to several programming languages, and
by opening new tabs, or new windows one can run different codes without their
interacting with each other. Atom editor is designed for professional coders or anyone
who needs to work with codes, and there are many useful packages. One of these  is atom-beautify \footnote{https://atom.io/packages/atom-beautify}, which tidies the codes,  when  the code is selected and \textit{Ctrl+Alt+B} is pressed.

Alternatively, the code can be run in-line.  Access to the in-line
technique can alter programming quite dramatically. The Hydrogen package
provides a modern approach,  using Jupyter kernels in Atom editor. It allows user
to choose which codes (the whole file, a single line, or a selection)
will be run. The combination of Hydrogen and Atom editor creates a unique
tool for running code in-line and in real time when developing   scripts.  This is because it
can keep track of objects and rerun. The user can add several tabs (\textit{Ctrl+N}) with access to the same kernel, so they have access to the same objects and
functions that have already loaded in the executed kernel. However,  opening
new windows launches new kernels independent of  the ones already
loaded. When R is selected from Grammar menu, Hydrogen automatically recognize
the kernel. Select a chunk of codes  and press \textit{Cmd+Enter}  (\textit{Ctrl+Enter} in Win) to run  the code. R's kernel
can also be activated manually from Rbox's menu or through pressing \textit{Shift+Ctrl+K}. The result of the run can be presented in another window (\textit{Shift+Ctrl+O}). Hydrogen can open small windows, called watch,  in which the user can type and run code without interfering with the main code (\textit{Shift+Ctrl+W}) and keep track of variables, which helps provide  great insight  into  value of variables. Rbox provides an interactive tool for coding that offers an effective method for debugging;  it is also an  appropriate environment in which  to run a graph.   These abilities are very powerful tools for developing codes.

\section{Conclusion}
\label{concluding} 
A considerable amount of research has been done to create a milestone statistical software, R offers an enhanced ease of use that has received  increasing attention from algorithm developers without  degrees in computer science. Supplementing article with well-written code helps researchers  find new applications for such tools  in applied sciences
where they have access to large amount of data. Because researchers are able to discover  novel
applications for the methods, more contact between method developers and
users helps us to draw better understandings.

Modern researchers in statistics must use R in conjunction with other software; hence,
access to an advanced editor is necessary,  and Atom editor provides such a tool. Here, we present a package for running R in this editor. Atom editor and its packages
provide a modern method of running codes; the codes can be run using access
terminals in Atom, or it can be executed in-line, providing  instant feedback on how users' data is structured.  This design provides an exploratory tool for developing code at different levels of detail and plays a very important role when coupled with testing and validating code in different
languages. This author assembled a series of codes to provide the best
possible interactive environment for R users carrying out projects with a modern IDE, whether the projects are scientific or not. In short,
this packages provides a high standard package for conducting research
and developing codes.

This package will be kept updated,  and  further investigation and improvement are planned to its performance. Unlike in Rstudio, the user can see almost all of  the code in Atom and its packages and change it.  R users are encouraged to participate and leave comments to aid in the development of  this package for R users.

\begin{table}[h]
\centering
\caption{Shortcut keys under Rbox}
\label{shorcut}
\begin{tabular}{lcc}
\hline
&\multicolumn{2}{c}{Shortcut keys}             \\ \cline{2-3}
Action     & macOS & Win \\ \hline
Select Grammar&Shift+Ctrl+L& Shift+Ctrl+L\\
Select Kernel &Shift+Ctrl+K&Shift+Ctrl+K\\
Run code &Cmd+Enter&Ctrl+Enter\\
Add Watch &Shift+Ctrl+W&Shift+Ctrl+W\\
Remove Watch &Shift+Ctrl+E&Shift+Ctrl+E\\
Show the result in new window&Shift+Ctrl+O& Shift+Ctrl+O\\
Run code in-line &Ctrl+Enter&Ctrl+Enter\\
Interrupt R &Ctrl+C&Ctrl+C\\
Quit or shutdown R  &Shift+Ctrl+Q&Shift+Ctrl+Q\\
Restart R  &Shift+Ctrl+R&Shift+Ctrl+R\\
Paste scripts in Terminal&  Alt+R&  Alt+R\\
Sort the codes&Ctrl+Alt+B&Ctrl+Alt+B\\\hline
\end{tabular}
\end{table}

\subsection*{Acknowledgments}
Author is  grateful to Leila Alimehr who has been very supportive in doing this project from beginning.

\end{document}